\documentclass[prl,twocolumn,showpacs,aps]{revtex4-1}

\usepackage{amsmath, amsfonts, amssymb, mathrsfs}
\usepackage{txfonts}
\usepackage{graphicx}
\usepackage{epsfig}
\usepackage{enumerate}
\usepackage{caption}
\usepackage{subcaption}

\usepackage{slashed}
\usepackage{ulem}

\newcommand{\bea}{\begin{eqnarray}}
\newcommand{\eea}{\end{eqnarray}}

\newcommand{\be}{\begin{equation}}
\newcommand{\ee}{\end{equation}}

\usepackage{ifpdf}
\ifpdf
\usepackage{epstopdf}
\fi

\usepackage[unicode]{hyperref}
\hypersetup{
	pdftitle={Schwarzian Theory and Cosmological Constant Problem},
	pdfauthor={Nian},
	pdfsubject={Schwarzian Theory and Cosmological Constant Problem},
	colorlinks=true,
	linkcolor=blue,
	citecolor=blue,
	filecolor=black,
	urlcolor=blue
}

\def\url#1{}

\makeatletter
\newcommand{\vast}{\bBigg@{4}}
\newcommand{\Vast}{\bBigg@{5}}
\makeatother

\usepackage{bm}
\usepackage{color}

\begin{document}

\title{Schwarzian Theory and Cosmological Constant Problem}
\author{Jun Nian}
\email{nianjun@ucas.ac.cn}
\affiliation{International Centre for Theoretical Physics Asia-Pacific, University of Chinese Academy of Sciences, 100190 Beijing, China}

\begin{abstract}
Observational data in cosmology indicate a small, positive, and nonvanishing cosmological constant that dominates the energy budget of the present universe. The origin of the cosmological constant from a quantum perspective remains unresolved, with a discrepancy of approximately 120 orders of magnitude between its observed value and theoretical estimates. Motivated by earlier work of Gibbons, we analyze the cosmological constant problem within a quantum-gravitational framework based on Schwarzian theory and its ensemble averaging. We then derive the phenomenological value of the dark energy density and obtain the corresponding equation of state. In this model, the cosmological constant arises from the ensemble average of time-reparametrization modes.
\end{abstract}

\maketitle

\textit{\bf Introduction.---} Over the past three decades, observational evidence has established that the expansion of the universe is accelerating \cite{SupernovaSearchTeam:1998fmf, SupernovaCosmologyProject:1998vns}. In the standard cosmological framework, this behavior is described by a small, positive cosmological constant that is approximately homogeneous on large scales and commonly referred to as dark energy. However, quantum field theoretic estimates of the vacuum energy density exceed the observed dark energy density by approximately 120 orders of magnitude, constituting the cosmological constant problem.

The quantum origin of dark energy has been investigated extensively, including prior to its observational confirmation. Numerous proposals have been put forward, making the cosmological constant problem a central topic in theoretical physics. Representative approaches include quantum field theoretic estimates \cite{Zeldovich:1967gd, Cohen:1998zx}, supersymmetric models \cite{WITTEN:1982, Cremmer:1983bf}, dynamical dark energy models such as quintessence, phantom, and quintom scenarios \cite{Peebles:1987ek, WETTERICH:1988, Zlatev:1998tr, Armendariz-Picon:2000nqq, Caldwell:1999ew, Feng:2004ad}, anthropic arguments \cite{Weinberg:1987dv}, quantum cosmology and baby-universe mechanisms \cite{HAWKING:1984, Coleman:1988tj}, string and membrane constructions \cite{Bousso:2000xa, Kachru:2003aw}, extra-dimensional models \cite{Rubakov:1983bz, Arkani-Hamed:2000hpr}, holographic dark energy \cite{Horava:2000tb, Li:2004rb}, and modified gravity frameworks \cite{Dvali:2003rk, Starobinsky:2007hu}. Reviews of these developments can be found in \cite{Weinberg:1988cp, Carroll:2000fy, Peebles:2002gy, Padmanabhan:2002ji}. Despite sustained effort, a framework that simultaneously accounts for observational constraints and theoretical consistency has not been established.

The cosmological constant becomes dynamically significant only when the matter density is sufficiently low, corresponding to the late stage of the matter-dominated epoch. During this period, matter in the observable universe is nonrelativistic and dilute. Gibbons and collaborators have shown that this regime can be described effectively within a Newtonian cosmological framework \cite{Duval:1990hj, Gibbons:2003rv, Ellis:2013xjx}. In addition, time reparametrizations within this framework can modify the effective cosmological constant \cite{Gibbons:2014zla}.

Motivated by earlier work of Gibbons and recent developments in quantum Schwarzian theory \cite{Stanford:2017thb, Mertens:2017mtv}, we study the cosmological constant problem within a quantum Schwarzian theory. We consider a $\Lambda$CDM-type model with vanishing cosmological constant and introduce time reparametrizations. Quantizing the time reparametrization modes through a Schwarzian path integral and performing the ensemble average at fixed, redshift-independent temperature yield a nonzero average cosmological constant. The resulting value reproduces the phenomenological dark energy density and the corresponding equation of state.

\textit{\bf Gibbons' Approach and Extension.---} Dark energy becomes dynamically significant near the end of the matter-dominated epoch. To examine the cosmological constant problem in this regime, we consider an effective description in the weak-field, nonrelativistic dust limit. In this limit, standard FLRW cosmology admits a Newtonian description \cite{Duval:1990hj, Gibbons:2003rv, Ellis:2013xjx}, which reproduces the Friedmann equations during matter domination and the scaling of the nonrelativistic matter density, $\rho \propto a^{-3}$. We adopt Newtonian cosmology as the background and study time reparametrizations relative to it.

We briefly review the relevant aspects of Newtonian cosmology (see the supplemental material for details). In the weak-field, nonrelativistic dust limit, the matter content can be approximated by the Newtonian action of $N$ particles with large but finite $N$,
\be
  S_N = \int dt \left[\sum_{1\leq i\leq N} \frac{1}{2} m_i \dot{{\bf r}}_i^2 - U ({\bf r}_i,\, t) \right]\, ,
\ee
where the dot ($\dot{\phantom{a}}$) denotes differentiation with respect to $t$. We introduce a time reparametrization accompanied by a rescaling of the spatial coordinates,
\be\label{eq:Time reparametrization - 1}
  t = f (\tilde{t})\, ,\quad {\bf r}_i = \tilde{{\bf r}}_i \sqrt{f' (\tilde{t})}\, ,
\ee
where $\tilde{t}$ is the new time variable and $f (\tilde{t})$ is a smooth function. The prime ($'$) denotes differentiation with respect to $\tilde{t}$. Requiring that the action retain its form in terms of $\tilde{t}$ and $\tilde{{\bf r}}_i$:
\be\label{eq:New action}
  S_N = \int d\tilde{t} \left[\sum_{1\leq i\leq N} \frac{1}{2} m_i (\tilde{\bf r}'_i)^2 - \widetilde{U} (\tilde{{\bf r}}_i,\, \tilde{t}) \right]\, ,
\ee
fixes the transformed potential to
\be\label{eq:New potential}
  \widetilde{U} = f'\, U \left(\tilde{{\bf r}}_i \sqrt{f'},\, f(\tilde{t}) \right) + \frac{1}{4} \{f,\, \tilde{t} \} \sum_{1 \leq i \leq N} m_i\, \tilde{{\bf r}}_i^2\, ,
\ee
where the Schwarzian derivative is defined by
\be
  \{f (\tilde{t}),\, \tilde{t} \} \equiv \frac{f''' (\tilde{t})}{f' (\tilde{t})} - \frac{3}{2} \left(\frac{f'' (\tilde{t})}{f' (\tilde{t})} \right)^2\, .
\ee
In Newtonian cosmology \cite{Duval:1990hj, Gibbons:2003rv, Ellis:2013xjx}, the second term in Eq.~\eqref{eq:New potential} plays the role of a cosmological constant. Time reparametrizations therefore generate a Schwarzian contribution to the potential, effectively modifying the cosmological constant \cite{Gibbons:2014zla}.

As the universe transitions from matter domination to dark energy domination, it may be approximated as a dilute system of nonrelativistic particles. In this regime, the kinetic term and the original Newtonian potential are subleading compared to the Schwarzian contribution in Eq.~\eqref{eq:New potential}. The action during this transition therefore reduces to
\begin{align}
  S_N & \approx \int d\tilde{t}\, \left[- \frac{1}{4} \{f,\, \tilde{t} \} \sum_{1 \leq i \leq N} m_i\, \tilde{{\bf r}}_i^2 \right] \nonumber\\
  {} & = - \frac{1}{4} \left(\sum_{1 \leq i \leq N} m_i\, \tilde{{\bf r}}_i^2 \right) \int d\tilde{t}\, \{f,\, \tilde{t} \} \nonumber\\
  {} & \equiv - C \int d\tilde{t}\, \{f,\, \tilde{t} \}\, ,
\end{align}
which is the standard one-dimensional Schwarzian action. The Schwarzian coupling in this case is
\be\label{eq:Define C}
  C \equiv \frac{1}{4} \left(\sum_{1 \leq i \leq N} m_i\, \tilde{{\bf r}}_i^2 \right)\, ,
\ee
corresponding to a moment-of-inertia-type quantity that is large for the present observable universe. This expression arises within the nonrelativistic approximation and reflects the dependence of the effective coupling on the global matter distribution.

We now apply time reparametrizations to the standard $\Lambda$CDM cosmology governed by the Friedmann equations and the continuity equation,
\begin{align}
  \frac{\ddot{a}}{a} & =	 - \frac{4 \pi G}{3} (\rho + 3 P) + \frac{\Lambda}{3}\, ,\label{eq:Friedmann-1}\\
  \left(\frac{\dot{a}}{a} \right)^2 & = \frac{8 \pi G}{3} \rho - \frac{k}{a^2} + \frac{\Lambda}{3}\, ,\label{eq:Friedmann-2}\\
  \dot{\rho} & = - \frac{3 \dot{a}}{a} (\rho + P)\, .\label{eq:Continuity}
\end{align}
Under the time reparametrization and rescaling
\be\label{eq:Time reparametrization - 2}
  t = f(\tilde{t})\, ,\quad a(t) = \tilde{a} (\tilde{t}) \sqrt{f'(\tilde{t})}\, ,
\ee
Eq.~\eqref{eq:Friedmann-1} and the continuity equation retain their form,
\begin{align}
  \frac{\tilde{a}''}{\tilde{a}} & = - \frac{4 \pi \widetilde{G} \rho_0}{3 \tilde{a}^3} + \frac{\widetilde{\Lambda}}{3}\, ,\label{eq:New Friedmann-1}\\
  \tilde{\rho}' & = - \frac{3 \tilde{a}'}{\tilde{a}} + \tilde{\rho}\, ,
\end{align}
where
\be\label{eq:Gtilde and Lambdatilde}
  \widetilde{G} = G \sqrt{f'}\, ,\quad \widetilde{\Lambda} = (f')^2 \Lambda - \frac{3}{2} \{f,\, \tilde{t}\}\, .
\ee
The transformation of Eq.~\eqref{eq:Friedmann-2} will be discussed below. Here, $G$ and $\Lambda$ denote the original couplings, while $\widetilde{G}$ and $\widetilde{\Lambda}$ represent their new values under the time reparametrization. As noted in Ref.~\cite{Gibbons:2014zla}, the cosmological constant can therefore be shifted by an appropriate choice of $f$.

In Gibbons' approach, all time reparametrizations are treated on equal footing, and no preferred choice is specified. Here, we extend this construction by implementing an ensemble average within quantum Schwarzian theory. The cosmological constant is then identified with the ensemble-averaged value.

\textit{\bf Cosmological Constant as Schwarzian Average.---} The Schwarzian derivative was first studied systematically by H.~A.~Schwarz in the nineteenth century \cite{Schwarz:1890gesammelte}. It subsequently appeared in two-dimensional conformal field theory and has recently played a central role in the study of the one-dimensional SYK model and two-dimensional JT gravity. The classical Schwarzian theory in Euclidean time is defined by the action
\be
  S_\text{Schw}^E = - C \int_0^\beta d\tau\, \{f (\tau),\, \tau \}\, ,
\ee
where $C$ is the Schwarzian coupling constant. The Euclidean version of quantum Schwarzian theory is defined by the partition function
\be\label{eq:Schwarzian path integral}
  Z_E (\beta) = \int_\mathcal{M} \mathcal{D} f\, e^{- S^E_\text{Schw} [f]}\, ,\quad \mathcal{M} = \text{Diff} (S^1) / SL(2, \mathbb{R})\, ,
\ee
where the functional integral is taken over the coset $\mathcal M$. Correlation functions in Euclidean signature are defined by
\be
  \langle \mathcal{O}_1 \cdots \mathcal{O}_n \rangle_E = \frac{1}{Z_E} \int_\mathcal{M} \mathcal{D} f\, \mathcal{O}_1 \cdots \mathcal{O}_n\, e^{- S^E_\text{Schw} [f]}\, .
\ee

Performing a Wick rotation, $\tilde t = -i\tau$, one obtains the Minkowski formulation of the Schwarzian theory. Correlation functions are defined by
\be\label{eq:Schwarzian average}
  \langle \mathcal{O}_1 \cdots \mathcal{O}_n \rangle_M = \frac{1}{Z_M} \int \mathcal{D} f\, \mathcal{O}_1 \cdots \mathcal{O}_n\, e^{i\, S^M_\text{Schw} [f]}\, .
\ee
with $Z_M$ the corresponding partition function. In particular, the ensemble-averaged Schwarzian derivative in Minkowski signature can be obtained from its Euclidean counterpart \cite{Stanford:2017thb, Mertens:2017mtv}, yielding
\be
  \langle \{f, \tilde{t}\, \} \rangle_M = - 2 \pi^2 A^2 T^2 - \frac{3 A T}{2 C}\, ,
\ee
where $C$ denotes the Schwarzian coupling, and $A$ is a positive integer fixed by matching physical conditions (see the supplemental material for details). The Schwarzian modes constitute the lowest-energy sector of quantum gravity and can modify infrared behavior at low temperatures \cite{Iliesiu:2020qvm, Liu:2024gxr, Brown:2024ajk, Liu:2024qnh, Emparan:2025sao, Nian:2025oei, PandoZayas:2025snm, Cremonini:2025yqe, Gouteraux:2025exs, Kanargias:2025vul}.

We now apply quantum Schwarzian theory to the cosmological constant problem. Starting from the reparametrized Friedmann equation \eqref{eq:New Friedmann-1}, we perform an ensemble average over time reparametrizations by embedding the equation in the Schwarzian path integral. In this construction, the one-dimensional Schwarzian action $S_N$ derived from Newtonian mechanics is identified with the Minkowski Schwarzian action $S^M_\text{Schw}$ in the corresponding path integral. This averaging procedure is analogous to the Schwinger-Dyson formalism \cite{Dyson, Schwinger} and has recently been employed in related contexts \cite{Jiang:2025cyl, Chen:2025rcc}.

Since only $\widetilde{G}$ and $\widetilde{\Lambda}$ in Eq.~\eqref{eq:New Friedmann-1} depend on the reparametrization function $f$ through Eq.~\eqref{eq:Gtilde and Lambdatilde}, the ensemble average reduces to averaging these couplings in the Schwarzian path integral, yielding
\be
  \frac{\tilde{a}''}{\tilde{a}} = - \frac{4 \pi\, \rho_0\, \langle\widetilde{G}\rangle_M}{3 \tilde{a}^3} + \frac{\langle\widetilde{\Lambda} \rangle_M}{3}\, ,
\ee
where $\langle\, \cdot\, \rangle_M$ denotes the expectation value in the Minkowski Schwarzian path integral. In the following, we assume Minkowski signature and suppress the subscript. We take the initial cosmological constant to vanish prior to reparametrization, $\Lambda = 0$, corresponding to a flat background. Eq.~\eqref{eq:Gtilde and Lambdatilde} then implies
\be
  \widetilde{\Lambda} = - \frac{3}{2} \{f, \tilde{t}\, \} \quad\Rightarrow\quad \langle\widetilde{\Lambda} \rangle = - \frac{3}{2} \langle\{f, \tilde{t}\, \} \rangle\, .
\ee
In the absence of a weighting functional, all reparametrizations would contribute equally. The Schwarzian path integral instead assigns nontrivial weights. The averaged Newton constant $\langle \widetilde{G} \rangle$ is obtained analogously. With these averaged couplings, the second Friedmann equation \eqref{eq:Friedmann-2} retains its form after ensemble averaging,
\be\label{eq:New Friedmann-2}
  \left(\frac{\tilde{a}'}{\tilde{a}} \right)^2 = \frac{8 \pi\, \rho_0\, \langle\widetilde{G}\rangle}{3 \tilde{a}^3} - \frac{k}{\tilde{a}^2} + \frac{\langle\widetilde{\Lambda}\rangle}{3}\, .
\ee

We identify the observed Newton constant and cosmological constant with the ensemble-averaged couplings $\langle\widetilde{G}\rangle$ and $\langle\widetilde{\Lambda}\rangle$, respectively. The corresponding averaged dark energy density is therefore
\be\label{eq:AveDarkEnergyDensity}
  \langle \rho_\Lambda \rangle = \frac{\langle \widetilde{\Lambda} \rangle}{8 \pi \langle \widetilde{G} \rangle} = - \frac{3}{16 \pi \langle \widetilde{G} \rangle}\, \langle \{f, \tilde{t}\, \} \rangle = \frac{3 \pi A^2 T^2}{8 \langle \widetilde{G} \rangle} + \frac{9 A T}{32 \pi C \langle \widetilde{G} \rangle}\, .
\ee
The first term arises from the classical saddle point $f_0 = \text{tanh} (\pi A T \tilde{t})$, while the second term represents the one-loop correction. Since $\langle \widetilde{G} \rangle$ corresponds to the measured value of Newton constant, we denote it by $G_N$ in the following.

\textit{\bf The Choice of Temperature.---} A central question is the choice of temperature in Eq.~\eqref{eq:AveDarkEnergyDensity}. Given the homogeneity of the dark energy density, the corresponding temperature should be universal across the observable universe.

A natural candidate is the de Sitter (dS) horizon temperature, defined as the Hawking temperature associated with the cosmological horizon,
\be\label{eq:T_dS}
  T_{dS} (z) = \frac{1}{2 \pi R (z)} = \frac{H (z)}{2 \pi}\, ,
\ee
where $R(z)$ is the horizon radius, which is related to the Hubble parameter as $R(z) = H(z)^{-1}$. A limitation of this choice is its time- or redshift-dependence. This reflects the fact that the universe contains matter and radiation and therefore does not correspond to an exact de Sitter spacetime.

For a pure de Sitter spacetime,
\be\label{eq:Lambda-dominant H}
  H = \sqrt{\frac{\Lambda}{3}}\, ,
\ee
as follows from Eq.~\eqref{eq:Friedmann-2}. Hence, we define $H_\Lambda \equiv \sqrt{\Lambda / 3}$ as the characteristic scale of the late-time, $\Lambda$-dominated universe \cite{Jazayeri:2016jav}. We will show below that the second term in Eq.~\eqref{eq:AveDarkEnergyDensity} is negligible for the present observable universe. Combining Eqs.~\eqref{eq:AveDarkEnergyDensity}, \eqref{eq:T_dS}, and \eqref{eq:Lambda-dominant H} then fixes the constant $A=2$. The averaged dark energy density therefore becomes
\be\label{eq:AveDarkEnergyDensity_New}
  \langle \rho_\Lambda \rangle = \frac{\langle \widetilde{\Lambda} \rangle}{8 \pi G_N} = - \frac{3}{16 \pi G_N}\, \langle \{f, \tilde{t}\, \} \rangle = \frac{3 \pi T^2}{2 G_N} + \frac{9 T}{16 \pi G_N C}\, .
\ee

Since our construction is based on averaging over time reparametrizations, the framework does not privilege any particular time slicing. A time-dependent de Sitter horizon temperature is therefore not suitable. Instead, we introduce a universal, time-independent temperature \cite{Gibbons:1977mu, Spradlin:2001pw, Bousso:2002ju, Cai:2005ra, Viaggiu:2016hth}:
\be\label{eq:T_Lambda initial}
  T_\Lambda \equiv \frac{H_\Lambda}{2 \pi} = \frac{H_0 \sqrt{\Omega_\Lambda^0}}{2 \pi}\, ,
\ee
where $H_\Lambda$ is defined in Eq.~\eqref{eq:Lambda-dominant H}. We assume that dark energy and matter are not in thermal equilibrium. The temperature $T_\Lambda$ isolates the $\Lambda$-dependent contribution and coincides with that of a pure de Sitter spacetime. It may be interpreted as the temperature measured by a freely falling observer at the center of the static patch of a pure de Sitter spacetime.

\begin{figure}[!htb]
  \begin{center}
     \includegraphics[width=0.48\textwidth]{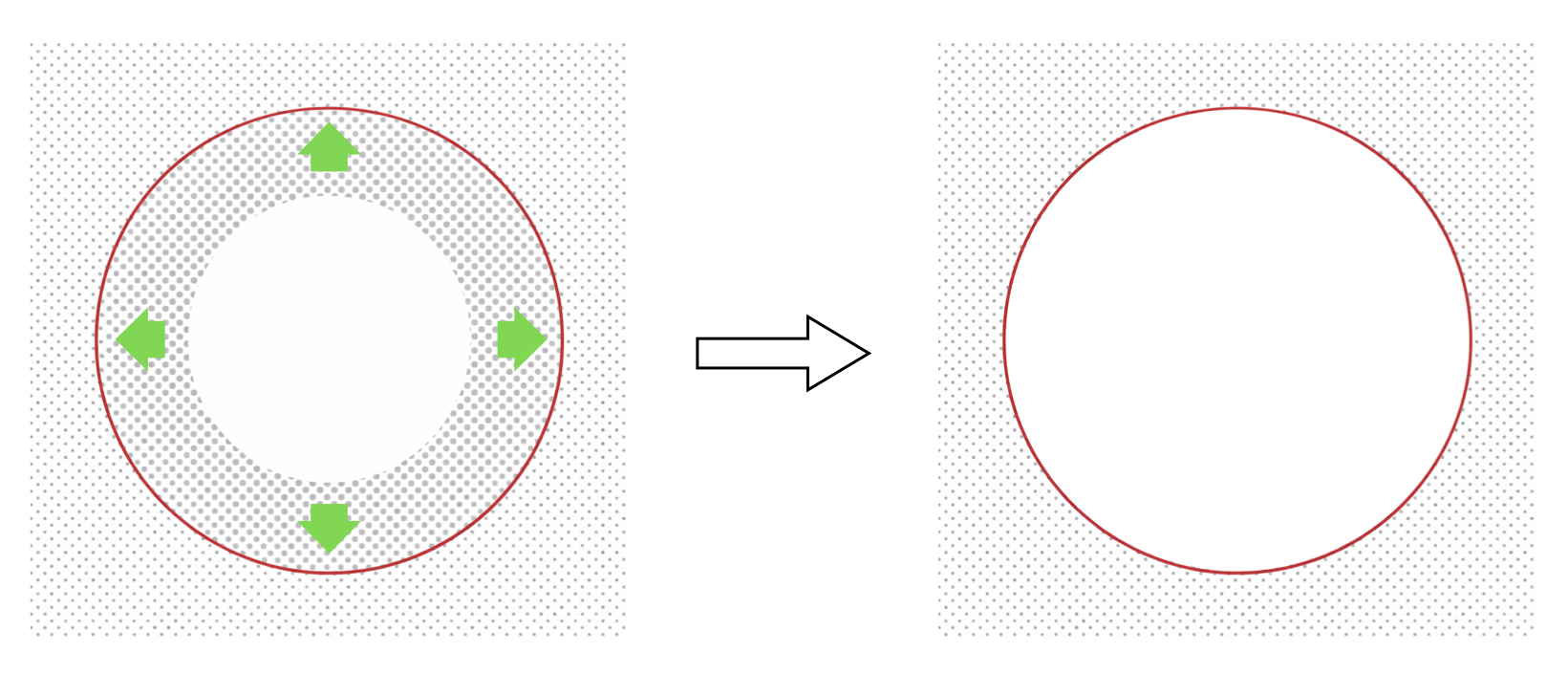}
     \caption{The physical apparent horizon $H(z)^{-1}$ eventually approaches the reference apparent horizon $R_\Lambda$ (the red circle).}
     \label{fig:Horizon-1}
  \end{center}
\end{figure}

Combining Eqs.~\eqref{eq:T_dS} and \eqref{eq:T_Lambda} yields another equivalent expression:
\begin{align}
  T_\Lambda & = T_{dS}(z)\, \frac{H_0 \sqrt{\Omega_\Lambda^0}}{H(z)} \nonumber\\
  {} & = \frac{T_{dS}(z)}{\sqrt{1 + \frac{\Omega_m^0}{\Omega_\Lambda^0} (1+z)^3 + \frac{\Omega_r^0}{\Omega_\Lambda^0} (1+z)^4 + \frac{\Omega_k^0}{\Omega_\Lambda^0} (1+z)^2}} \nonumber\\
  {} & = \frac{T_{dS}(z)}{\sqrt{1 + \frac{\Omega_m (z)}{\Omega_\Lambda (z)} + \frac{\Omega_r (z)}{\Omega_\Lambda (z)} + \frac{\Omega_k (z)}{\Omega_\Lambda (z)}}}\, ,\label{eq:T_Lambda}
\end{align}
which can be measured at any redshift $z$ but is independent of time or redshift. This temperature admits an alternative interpretation. The time-independent $\Lambda$-normalized temperature $T_\Lambda$ corresponds to a reference apparent horizon that can be defined at the onset of radiation domination, with radius
\be\label{eq:would-be horizon}
  R_\Lambda = \frac{1}{H_\Lambda} = \frac{1}{H_0 \sqrt{\Omega_\Lambda^0}} > \frac{1}{H_0}\, .
\ee
The physical apparent horizon remains smaller than this reference scale for most of cosmic history and becomes comparable to it only at late times, $z \in [0,\, 2]$. In the asymptotic future, the apparent horizon approaches $R_\Lambda$, and the spacetime approaches pure de Sitter geometry (see Fig.~\ref{fig:Horizon-1}). Thus, $R_\Lambda = 1 / (2 \pi T_\Lambda)$ sets the maximal radius of the physical apparent horizon. Although $T_\Lambda$ can be fixed from observational data, a first-principles derivation is not presently available; instead, we interpret it as determined by the maximal size of the observable universe in the asymptotic future. As long as the matter contribution to the cosmic energy budget remains non-negligible relative to dark energy, the apparent horizon radius continues to increase, qualitatively accounting for the smallness of $H_\Lambda$, $T_\Lambda$, and the observed cosmological constant.

\textit{\bf New Theoretical Cosmological Constant.---} From Eq.~\eqref{eq:Define C}, the present value of $C$ for the observable universe is $3.969\times 10^{163}\, \text{GeV}^{-1}$. Substituting this value into Eq.~\eqref{eq:AveDarkEnergyDensity_New} shows that the second term is negligible relative to the first. The ensemble average is therefore dominated by a single saddle configuration $f_0$, corresponding to a classical solution of the Schwarzian action. In the following, we retain only the leading term in Eq.~\eqref{eq:AveDarkEnergyDensity_New} when evaluating numerical predictions for the present universe.

Finally, using the time-independent $\Lambda$-normalized temperature $T_\Lambda$, \eqref{eq:AveDarkEnergyDensity_New} yields the present averaged dark energy density,
\be\label{eq:AveDarkEnergyDensity_Final}
  \langle \rho_\Lambda \rangle \approx \frac{3 \pi T_\Lambda^2}{2 G_N} = \frac{3 H_0^2\, \Omega_\Lambda^0}{8 \pi G_N} \approx 2.62\times 10^{-47}\, \text{GeV}^4\, ,
\ee
in agreement with observations. The resulting expression coincides with the standard phenomenological form.

Since the averaged dark energy density is time-independent in this model, $\langle \rho_\Lambda \rangle \propto a^0$, the equation of state coincides with that of a cosmological constant,
\be
  P = \omega \rho \quad\text{with}\quad \omega = -1\, .
\ee

Within this framework, the predicted dark energy density and equation of state are consistent with current observational constraints on constant-$\omega$ models in the range $z \in [0,\, 2]$ \cite{Planck:2018vyg, Brout:2022vxf, DESI:2024mwx}. In this construction, the cosmological constant arises from the ensemble average over time reparametrizations, corresponding to the lowest-lying modes of quantum gravity.

\textit{\bf The Fate of Universe.---} The second term in Eq.~\eqref{eq:AveDarkEnergyDensity_New}, neglected thus far, evaluates at the $\Lambda$-normalized temperature to
\be
  \frac{9 T_\Lambda}{16 \pi G_N C} = \frac{9 H_0 \sqrt{\Omega_\Lambda^0}}{32 \pi^2 G_N C}\, .
\ee
Although strongly suppressed by the present large value of $C$, this contribution grows with time and eventually dominates in the asymptotic future, as shown below.

The parameter $C$, defined in Eq.~\eqref{eq:Define C}, corresponds to the moment-of-inertia-type quantity evaluated within the cosmological event horizon. Strictly speaking, $C$ is time-dependent, and only its present contribution to the dark energy density is negligible. In the asymptotic future, the cosmological event horizon approaches the reference apparent horizon given in Eq.~\eqref{eq:would-be horizon}. Owing to the exponential growth of the scale factor, however, the total matter mass enclosed within the event horizon decreases. Consequently, the effective moment of inertia, and hence $C$, tends toward zero at late times. If the time-independent temperature $T_\Lambda$ is retained in Eq.~\eqref{eq:AveDarkEnergyDensity_New}, the second term eventually dominates, leading to a time-dependent dark energy density that increases with time.

\begin{figure}[!htb]
  \begin{center}
     \includegraphics[width=0.48\textwidth]{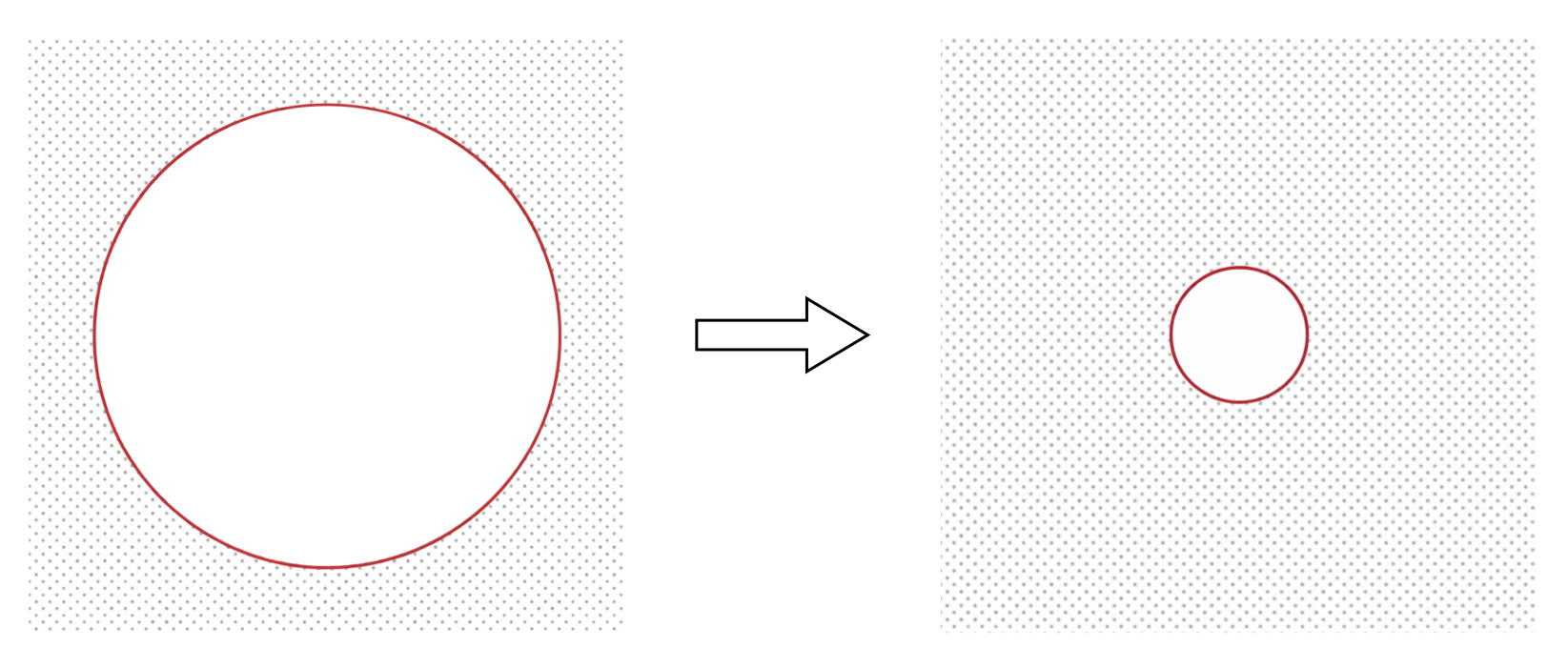}
     \caption{In the $\Lambda$-dominated universe, the physical and the reference apparent horizons coincide and eventually both shrink in the far future.}
     \label{fig:Horizon-2}
  \end{center}
\end{figure}

From the scaling relation $\rho \propto a^{-3 (1 + \omega)}$, the equation-of-state parameter for dark energy in this model evolves from the present value $\omega_\Lambda = -1$ toward $\omega_\Lambda < -1$ at late times. The model therefore predicts a future phantom-like phase that can culminate in a Big Rip singularity (see Fig.~\ref{fig:Horizon-2}), barring additional low-energy phase transitions. Consequently, $T_\Lambda$ does not remain constant at very late times but instead begins to increase, causing $\langle \rho_\Lambda \rangle$ to grow more rapidly. In this sense, the initial value given by Eq.~\eqref{eq:T_Lambda initial} represents a lower bound for $T_\Lambda$.

\textit{\bf Discussion.---} In this letter, we revisit and extend earlier work by Gibbons. Combining established results with recent developments, we interpret the observed cosmological constant as the Schwarzian ensemble average over time reparametrizations, with the Schwarzian action emerging from Newtonian cosmology near the end of the matter-dominated epoch. Since Newtonian cosmology provides an effective one-dimensional description of time evolution, it offers a natural connection to the one-dimensional Schwarzian theory of time reparametrizations. Within this framework, the dark energy density and equation of state follow directly.

We note that Ref.~\cite{Gibbons:2014zla} also considered an alternative time rescaling that induces a conformal transformation of the metric, distinct from Eqs.~\eqref{eq:Time reparametrization - 1} and \eqref{eq:Time reparametrization - 2} adopted here. The reparametrizations employed in this letter preserve the form of the equations of motion and are therefore natural from the perspective of a fixed local observer.

We also note related work in the literature. Ref.~\cite{Lee:2008vn} investigated the possibility that dark energy arises from cosmic Hawking radiation. However, the resulting dark energy density and equation of state agree with observations only approximately and not exactly. Gibbons' construction is also closely related to the approach of Henneaux and Teitelboim \cite{Henneaux:1989zc}, who connected cosmic time and the cosmological constant through an analysis of a constrained Hamiltonian system. The present framework, therefore, suggests a broader connection to the problem of time, which has been examined in recent studies \cite{Bojowald:2010xp, Klinger:2025hjp}.

Further theoretical developments and observational tests of the framework are warranted. Recent DESI data suggest possible deviations from a strictly constant $\Lambda$, highlighting the importance of distinguishing among competing models. More precise measurements, particularly of $H(z)$, $\rho_\Lambda (z)$, and $\omega_\Lambda (z)$ at low redshifts, will be essential for stringent tests of the present predictions.

The averaging of time reparametrizations proposed here may also be extended to the inflationary epoch. A consistent implementation near the end of inflation, however, requires a more refined theoretical treatment. We defer this analysis to future work.

\section*{acknowledgments}

The author would like to thank Jun-Peng Cao, Xiao-Yong Chu, Li Li, Yun-Song Piao, Zhao-Long Wang, and Cong-Yuan Yue for helpful discussions. This work is supported in part by the NSFC under grants No.~12375067 and No.~12547104. The author also thanks Hangzhou Institute for Advanced Study, Xiamen University, and Northwest University for their warm hospitality during the final stage of this work.

\bibliography{SchwarzLambda}

\newpage

\appendix

\section{Review of Newtonian Cosmology}

In this appendix, we briefly review the Newtonian cosmology by Gibbons as a homogeneous, low-velocity, weak-field, dust limit of the standard relativistic cosmology on the FLRW metric \cite{Duval:1990hj, Gibbons:2003rv, Ellis:2013xjx}.

The Newtonian cosmology can be derived from the standard Newtonian action of $N$ particles:
\be\label{app eq:action}
  S = \int dt\, \left[\sum_{i=1}^N \frac{1}{2} m_i \dot{{\bf r}}_i^2 + \frac{1}{2} \sum_{i \neq j} \frac{G m_i m_j}{|{\bf r}_i - {\bf r}_j|} \right]\, .
\ee
Imposing the scale factor $a(t)$ and the comoving coordinates ${\bf x}_i$ via
\be
  {\bf r}_i (t) = a(t)\, {\bf x}_i\, ,
\ee
we can rewrite the action \eqref{app eq:action} as
\be\label{app eq:action 2}
  S[a] = \int dt\, \left(\frac{1}{2} I \dot{a}^2 + \frac{V}{a} \right)\, ,
\ee
where
\be
  I \equiv \sum_{i=1}^N m_i |{\bf x}_i|^2\, ,\quad V \equiv \frac{1}{2} \sum_{i \neq j} \frac{G m_i m_j}{|{\bf x}_i - {\bf x}_j|}\, .
\ee
Adding a cosmological constant $\Lambda$ to the action \eqref{app eq:action 2}, we obtain the final action of the Newtonian cosmology:
\be\label{app eq:action final}
  S[a] = \int dt\, \left(\frac{1}{2} I \dot{a}^2 + \frac{V}{a} + \frac{\Lambda}{6} I a^2 \right)\, .
\ee
This action essentially defines a one-dimensional effective theory for the homogeneous, low-velocity, weak-field, dust limit of the matter-dominant universe with the cosmological constant.

The conservation of energy from the action \eqref{app eq:action final} is
\begin{align}
  {} & \frac{1}{2} I \dot{a}^2 - \frac{V}{a} - \frac{\Lambda}{3} I a^2 = E \\
  \Rightarrow\quad & \frac{\dot{a}^2}{a^2} = \frac{2 V}{I a^3} + \frac{\Lambda}{3} + \frac{2 E}{I a^2}\, ,
\end{align}
which can be exactly identified with one of the Friedmann equations for dust with the cosmological constant:
\be
  \frac{\dot{a}^2}{a^2} = \frac{8 \pi G}{3} \frac{\rho_0}{a^3} + \frac{\Lambda}{3} - \frac{k}{a^2}\, .
\ee
We can also derive the equation of motion from \eqref{app eq:action final}:
\be
  \frac{\ddot{a}}{a} = - \frac{V}{I a^3} + \frac{\Lambda}{3}\, ,
\ee
which exactly matches the other Friedmann equation for dust with the cosmological constant:
\be
  \frac{\ddot{a}}{a} = - \frac{4 \pi G}{3} \frac{\rho_0}{a^3} + \frac{\Lambda}{3}\, .
\ee

In summary, the Newtonian cosmology is a legitimate weak-field, dust limit of the relativistic FLRW cosmology. It provides a Newtonian description of the matter-dominant universe with a cosmological constant. Therefore, we can use it as an effective background for weak-field, sub-horizon, low-energy quantum-gravity corrections. However, as a background-level effective theory, it does not capture global geometry (e.g., horizon structure), super-horizon modes, and relativistic corrections.

\section{Quantum Average in Schwarzian Theory}

In this appendix, we collect some known facts from the literature on Schwarzian theory and derive new results relevant to this work.

We have introduced the cosmic time $\tilde{t}$ in the Minkowski signature in the main text. It can be related to the Euclidean time $\tau$ by a Wick rotation:
\be
  \tilde{t} = - i \tau\, .
\ee
In the literature, the classical Schwarzian theory with the Euclidean time $\tau$ is defined by the action:
\be\label{eq:Schwarzian action Eucl}
  S_{\text{Schw}}^E [f] = - C \int_0^\beta d\tau\, \{f(\tau),\, \tau \}_E\, ,
\ee
where $f(\tau)$ runs over the space of diffeomorphisms on the thermal circle, $\text{Diff} (S^1)$, and obeys the periodic boundary condition $f(\tau + \beta) = f(\tau)$, which originates from the condition \cite{Stanford:2017thb, Mertens:2017mtv}:
\be\label{eq:f and phi}
  f(\tau) = \text{tan} \left(\frac{\pi\, \phi (\tau)}{\beta} \right)\quad \text{with}\quad \phi (\tau + \beta) = \phi (\tau) + \beta\, .
\ee

The Euclidean version of the Schwarzian derivative is defined as
\be\label{eq:Schwarzian derivative Eucl}
  \text{Sch} (\tau)_E \equiv \{f(\tau),\, \tau \}_E = \frac{d^3 f/ d\tau^3}{df / d\tau} - \frac{3}{2} \left(\frac{df^2 / d\tau^2}{df / d\tau} \right)^2\, .
\ee
The action $S_{\text{Schw}}^E [f]$ is invariant under the $SL(2, \mathbb{R})$ M\"obius transformations acting on $f$:
\be
  f \to \frac{a f + b}{c f + d}\, ,\quad a, b, c, d \in \mathbb{R}\,\, \text{with}\,\, a d - b c = 1\, .
\ee
The Euclidean version of quantum Schwarzian theory has the partition function:
\begin{align}
  Z_E (\beta) & = \int_\mathcal{M} df\, e^{- S^E_{\text{Schw}} [f]} \nonumber\\
  {} & = \int_\mathcal{M} df\, \text{exp} \left[C \int_0^\beta d\tau\, \{f(\tau),\, \tau \}_E\ \right]\, ,
\end{align}
where
\be
  \mathcal{M} = \text{Diff} (S^1) / SL(2, \mathbb{R})\, .
\ee

The Euclidean partition function $Z_E$ can be computed exactly using the fermionic localization technique \cite{Stanford:2017thb}, or using the two-dimensional Virasoro CFT method \cite{Mertens:2017mtv}. The result is
\be
  Z_E (\beta) = \left(\frac{2 \pi C}{\beta} \right)^{3/2} \text{exp} \left(\frac{2 \pi^2 C}{\beta} \right)\, .
\ee
Consequently, some Euclidean correlation functions with the Schwarzian derivatives as operators can be computed by differentiating the partition function with respect to $C$. For example:
\begin{align}
  \langle \text{Sch} (\tau) \rangle_E & \equiv \frac{1}{Z_E} \int \mathcal{D} f\, \text{Sch} (\tau)_E\, e^{- S^E_{\text{Schw}} [f]} \nonumber\\
  {} & = \frac{2 \pi^2}{\beta^2} + \frac{3}{2 \beta C}\, ,\\
  \int_0^\beta d\tau\, \langle \text{Sch} (\tau) \rangle_E & = \frac{2 \pi^2}{\beta} + \frac{3}{2 C}\, .	
\end{align}

In the main text, we have obtained a classical Schwarzian theory in the Minkowski signature from the Newtonian cosmology:
\be
  S_{\text{Schw}}^M [f] = - C \int d\tilde{t}\, \{f(\tilde{t}),\, \tilde{t} \}_M\, ,
\ee
where the Minkowski version of the Schwarzian derivative is defined as
\be\label{eq:Schwarzian derivative Mink}
  \text{Sch} (\tilde{t})_M \equiv \{f(\tilde{t}),\, \tilde{t} \}_M = \frac{d^3 f/ d\tilde{t}\,^3}{df / d\tilde{t}} - \frac{3}{2} \left(\frac{df^2 / d\tilde{t}\,^2}{df / d\tilde{t}} \right)^2\, .
\ee
The Schwarzian derivatives in two signatures, \eqref{eq:Schwarzian derivative Eucl} and \eqref{eq:Schwarzian derivative Mink}, are not equal. Instead, they are related in the following way:
\begin{align}
  {} & \{f(\tilde{t}),\, \tilde{t} \}_M = \text{Sch} (\tilde{t})_M \nonumber\\
  =\, & \frac{d^3 f/ d\tilde{t}\,^3}{df / d\tilde{t}} - \frac{3}{2} \left(\frac{df^2 / d\tilde{t}\,^2}{df / d\tilde{t}} \right)^2 = - \left[\frac{d^3 f/ d\tau^3}{df / d\tau} - \frac{3}{2} \left(\frac{df^2 / d\tau^2}{df / d\tau} \right)^2 \right] \nonumber\\
  =\, & - \{f(\tau),\, \tau \}_E = - \text{Sch} (\tau)_E\, .\label{eq:sign difference}
\end{align}
Consequently, the Minkowski partition function of the quantum Schwarzian theory can be related to the Euclidean one via:
\begin{align}
  Z_M & = \int df\, e^{i\, S^M_{\text{Schw}} [f]} \nonumber\\
  {} & = \int df\, \text{exp} \left[- i\, C \int d\tilde{t}\, \{f (\tilde{t}),\, \tilde{t} \}_M \right] \nonumber\\
  {} & = \int df\, \text{exp} \left[C \int d\tau\, \{f(\tau),\, \tau \}_E \right] \nonumber\\
  {} & = Z_E\, .
\end{align}
Similar to the Euclidean case, the Minkowski correlation function $\langle \text{Sch} (\tilde{t}) \rangle_M$ can also be obtained by differentiating $Z_M$ with respect to $C$. However, due to the overall sign difference \eqref{eq:sign difference}, the result is
\begin{align}
  \langle \text{Sch} (\tilde{t}) \rangle_M & \equiv \frac{1}{Z_M} \int \mathcal{D} f\, \text{Sch} (\tilde{t})_M\, e^{i\, S^M_{\text{Schw}} [f]} \nonumber\\
  {} & = - \langle \text{Sch} (\tau) \rangle_E \nonumber\\
  {} & = - \frac{2 \pi^2}{\beta^2} - \frac{3}{2 \beta C}\, ,
\end{align}
which is consistent with the known result in the literature \cite{Jensen:2016pah}.

There is still a subtlety that should be pointed out. For the Euclidean version of the Schwarzian theory, the time reparametrization mode $f$ is required to satisfy the periodic boundary condition $f(\tau + \beta) = f(\tau)$. In fact, this periodic boundary condition and the corresponding condition \eqref{eq:f and phi} allow an additional parameter $A \in \mathbb{Z}_+$ such that
\be
  f(\tau) = \text{tan} \left(\frac{\pi\, A\, \phi (\tau)}{\beta} \right)\quad \text{with}\quad \phi (\tau + \beta) = \phi (\tau) + \beta\, ,
\ee
which is also a classical solution to the equation of motion from the variation of the action \eqref{eq:Schwarzian action Eucl}. Effectively, the time reparametrization mode $f$ has a new period $\beta / A$, i.e.,
\be
  f \left(\tau + \frac{\beta}{A} \right) = f(\tau)\, ,\quad A \in \mathbb{Z}_+\, ,
\ee
under which the old periodic boundary condition $f(\tau + \beta) = f(\tau)$ is also satisfied.

Repeating the same computation steps, all the results are modified by this new parameter $A$. In particular, the ensemble average of the Schwarzian derivative in Minkowski quantum Schwarzian theory has an expression with the parameter $A$:
\be
  \langle \text{Sch} (\tilde{t}) \rangle_M = - \frac{2 \pi^2 A^2}{\beta^2} - \frac{3 A}{2 \beta C}\, ,\quad A \in \mathbb{Z}_+\, ,
\ee
where $A$ needs to be fixed by matching physical conditions.

\bibliographystyle{utphys}

\end{document}